\RequirePackage{fix-cm}
\documentclass[twocolumn]{svjour3}          
\smartqed  
\usepackage{graphicx}
\usepackage{xcolor}
\definecolor{darkgreen}{RGB}{0,150,0}
\definecolor{darkred}{RGB}{150,0,0}
\definecolor{dyellow}{RGB}{255,150,0}
\usepackage{amsmath,amssymb}
\usepackage{tikz}
\begin{document}

\title{Phase-field modelling for fatigue crack growth under laser-shock-peening-induced residual stresses
}


\author{Martha Seiler \and Sören Keller \and Nikolai Kashaev \and Benjamin Klusemann \and Markus Kästner }


\institute{M. Seiler and M. Kästner \at
              Chair of Computational and Experimental Solid Mechanics, TU Dresden, Germany \\
              \email{markus.kaestner@tu-dresden.de}           
           \and
          S. Keller and N. Kashaev and B. Klusemann\at
             Institute of Materials Research, Materials Mechanics,  Helmholtz-Zentrum Geesthacht, Germany
             \and 
          B. Klusemann \at
          Institute of Product and Process Innovation, Leuphana University of Lüneburg, Germany
}

\date{Received: date / Accepted: date}

\maketitle

\begin{abstract}
For the fatigue life of thin-walled components, not only fatigue crack initiation, but also crack growth is decisive. The phase-field method for fracture is a powerful tool to simulate arbitrary crack phenomena. Recently, it has been applied to fatigue fracture. Those models pose an alternative to classical fracture-mechanical approaches for fatigue life estimation. In the first part of this paper, the parameters of a phase-field fatigue model are calibrated and its predictions are compared to results of fatigue crack growth experiments of aluminium sheet material. In the second part, compressive residual stresses are introduced into the components with the help of laser shock peening. It is shown that those residual stresses influence the crack growth rate by retarding and accelerating the crack. In order to study these fatigue mechanisms numerically, a simple strategy to incorporate residual stresses in the phase-field fatigue model is presented and tested with experiments. The study shows that the approach can reproduce the effects of the residual stresses on the crack growth rate.
\keywords{Laser shock peening \and Fatigue crack growth \and Phase-field modelling \and Residual stresses}
\end{abstract}

\section{Introduction}
Fatigue fracture is one of the most common causes of component failure. Still, the underlying physical phenomena, especially on the microscale \cite{bathias_fatigue_2010}, are not fully understood. Simulating fatigue fracture, e.\,g. via the Finite Element Method (FEM), is often very time-con\-su\-ming, as several hundred to millions of load cycles have to be simulated. 
Generally, the fatigue life of a component until fracture can be divided into the crack initiation and the crack propagation stage. While crack initiation often takes the major part, in thin-walled specimen like fuselage shells, crack growth is crucial for the design process. The growth rate of the long, visible cracks determines the maintenance interval.

Particularly, residual  stresses created  by  the  process  of  laser  shock  peening (LSP) can  be used to deliberately influence the fatigue crack growth (FCG) rate of long cracks, as demonstrated for the aluminium alloy AA2024 \cite{hombergsmeier2014fatigue,tan2004laser}. The application of laser shock peening aims at the introduction of compressive residual stresses in regions susceptible to fatigue, where the process provides a relatively high penetration depth as well as surface quality \cite{peyre1996laser}. These compressive residual stresses interact with the applied stresses of the fatigue load cycles and reduce the fatigue crack driving quantity. However, compressive residual stresses are always accompanied by tensile residual stresses to meet stress equilibrium. While compressive residual stresses are expected to retard the fatigue crack growth, tensile residual stresses may lead to increased fatigue crack growth rates reducing the fatigue life of the component, as shown by \cite{keller_experimentally_2019}. Thus, the efficient application of residual stress modification techniques needs the precise prediction of the residual stress field to determine the FCG rate. 

In order to estimate the fatigue life of residual stress affected components, mainly two types of models are used. Empirical concepts based on Wöhler curves mostly evaluate the fatigue life until crack \textit{initiation} and often treat residual stresses as mean stresses \cite{landersheim_analyse_2009}. For the remaining fatigue life during crack \textit{growth}, models based on fracture mechanics are often used. In this context, residual stresses can be applied with the eigenstrain approach. The residual stress field is applied with the help of a fictitious temperature field \cite{benedetti_numerical_2010,keller_experimentally_2019} and considered with effective stress intensity factors in the fatigue crack growth computations.
In contrast to the previously mentioned approaches, in this contribution, an FEM framework is applied which covers both crack initiation and growth -- the phase-field method. However, the focus of this paper lies on the fatigue crack \textit{growth}.
 
The phase-field method has become a popular tool to simulate fracture phenomena because of its capability to treat arbitrary crack paths in a straight-forward way. This is possible due to a second field variable which describes the crack topology, making mesh alterations due to crack growth redundant. Originally formulated for static brittle fracture \cite{miehe_thermodynamically_2010} by simply regularising the Griffith criterion \cite{griffith_phenomena_1921} for crack growth, the phase-field method has now been applied to a large variety of materials and phenomena like e.\,g. ductile fracture \cite{miehe_phase_2016,ambati_phasefield_2015}.

However, the phase-field modelling of \textit{fatigue} fracture has been addressed only recently. While some models reduce the crack resistance of the material as a result of its cyclic degradation \cite{carrara_framework_2019,seiler_efficient_2020,mesgarnejad_phasefield_2018}, others increase the crack driving force \cite{schreiber_phase_2020,haveroth_nonisothermal_2020}. There are first approaches to cover plastic \cite{ulloa_phasefield_2019} and viscous \cite{loew_fatigue_2020} materials. To tackle the crucial problem of computational time when simulating repetitive loading, Loew et al. \cite{loew_accelerating_2020} and Schreiber et al. \cite{schreiber_phase_2020} use cycle jump techniques. Seiler et al. \cite{seiler_efficient_2020} approached the problem by incorporating a classic fatigue concept -- the local strain approach (LSA) -- into the model, enabling the simulation of several load cycles within only one increment. However, this model has been studied only qualitatively.

In this contribution, the phase-field fatigue model of \cite{seiler_efficient_2020} is calibrated and validated using FCG experiments of aluminium AA2024. Moreover, a straightforward strategy to include residual stresses in the model is presented. Additional FCG experiments are conducted with specimen  in which residual stresses are introduced deliberately with LSP. Those residual stresses are analysed with the incremental hole drilling method and serve as an initial state for the fatigue simulation.

The paper is structured as follows: In Section~\ref{sec:model}, the model formulation is recapitulated, including the underlying phase-field equations for brittle fracture and its extension to fatigue, and the incorporation of residual stresses is explained. Section~\ref{sec:Exp} deals with the LSP experiments for the creation of residual stresses, the experimental determination of the resulting residual stress state, as well as the FCG experiments. Section~\ref{sec:Sim} contains the numerical predictions of the proposed model including the model parameter calibration and the comparison to experimental results. The conclusion follows in Section~\ref{sec:Conc}.

\section{Model framework}
\label{sec:model}
The model used in this publication is extensively described and qualitatively studied in \cite{seiler_efficient_2020}. Therefore, the model formulation is only outlined briefly here, starting with the basis of the framework, the phase-field method for brittle fracture, as well as its extension to fatigue. Then, going one step further, the incorporation of residual stresses in the model is described in detail.

\subsection{Phase-field method for fracture}
\label{sec:model_PF}

\begin{figure}
    \def\svgwidth{\linewidth}  
    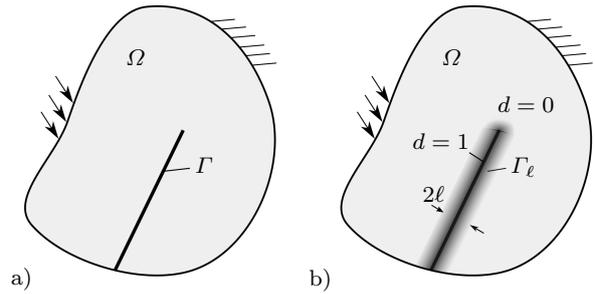
    \caption{Fractured domain $\Omega$ with crack surface $\Gamma$. \textbf{(a)} Sharp representation of crack topology. \textbf{(b)} Regularised representation: The crack is described by the phase-field variable $d=1$, while $d=0$ represents undamaged material. The crack is regularised over the length scale $\ell$.
    \label{fig:Model_PF}}
\end{figure}

\begin{figure*}
    \def\svgwidth{\textwidth}  
    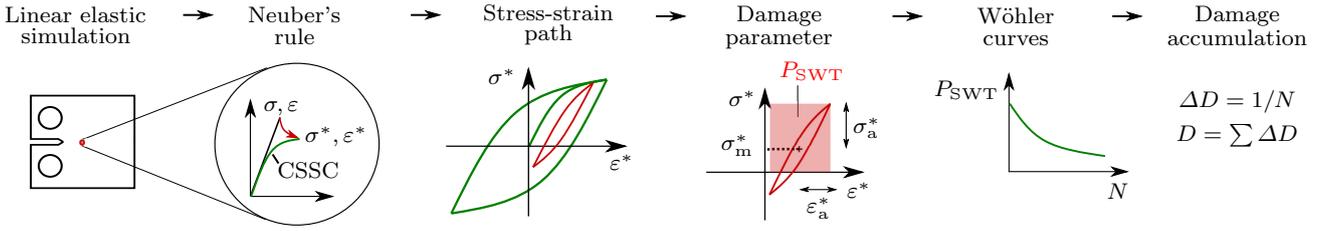
    \caption{Scheme of local strain approach (LSA), with which the fatigue life variable $D$ is determined at every material point. 
    }
    \label{fig:LSA}      
\end{figure*}

The phase-field method for brittle fracture is based on the Griffith criterion \cite{griffith_phenomena_1921} for crack growth, which requires the energy release rate to be equal to the critical energy release rate or fracture toughness $\mathcal{G}_\mathrm{c}$. This criterion was brought to a variational form in \cite{francfort_revisiting_1998} and regularised for convenient numerical implementation in  \cite{bourdin_variational_2008}.
During the regularisation, an additional field variable $d\in[0,1]$ is introduced. The diffuse description of the crack smoothly bridges the entirely intact ($d=0$) and totally broken ($d=1$) state. In this way, the crack topology can be described without any mesh modifications, allowing for a straightforward modelling of arbitrary crack paths. See Fig.~\ref{fig:Model_PF} for a graphical explanation of the regularisation. 

Using the regularisation length scale $\ell$, the regularised energy functional $\Pi_\ell$ in the domain $\Omega$ can be written as
\begin{equation}
    \Pi_\ell = \int_{\Omega} g(d)\,\psi^\mathrm{e}(\boldsymbol{\varepsilon}) \, \mathrm{d} V + \int_{\Omega} \mathcal{G}_\mathrm{c}{\frac{1}{2\ell}(d^2+\ell^2|\nabla d |^2)}\,\mathrm{d}V.
\end{equation}
In the small strain linear elastic setting. The elastic strain energy density is 
\begin{equation}
    \psi^\mathrm{e}=\frac{1}{2}\lambda\,\mathrm{tr}^2(\boldsymbol{\varepsilon})+\mu\,\mathrm{tr}(\boldsymbol{\varepsilon}^2) = \frac{1}{2}\,\boldsymbol{\varepsilon}: \mathbb{C} : \boldsymbol{\varepsilon}
\end{equation}
with the Lamé constants $\lambda$ and $\mu$, the elasticity tensor $\mathbb{C}$ and the strain
\begin{equation}
    \boldsymbol{\varepsilon} = \frac{1}{2}\left(\nabla \boldsymbol{u} + (\nabla \boldsymbol{u})^\top\right).
\end{equation}
The degradation function $g(d)=(1-d)^2$ models the loss of stiffness due to the developing crack, coupling displacement field $\boldsymbol{u}$ and phase-field $d$. Consequently, the stress is given by
\begin{equation}
\label{eq:stress}
    \boldsymbol{\sigma}=g(d)\frac{\partial\psi^\mathrm{e}}{\partial \boldsymbol{\varepsilon}} = g(d)\, \mathbb{C}:\boldsymbol{\varepsilon}.
\end{equation}
The governing equations of the coupled problem obtained from the variation $\delta\Pi_\ell=0$
\begin{equation} \label{eq:goveq}
    \boldsymbol{0}=\mathrm{div}\,\boldsymbol{\sigma} \qquad \qquad
    \mathcal{G}_\mathrm{c}\left(d-\ell^2\Delta d\right) = (1-d)2\ell\underbrace{\psi^\mathrm{e}(\boldsymbol{\varepsilon})}_\mathcal{H}
\end{equation}
are subject to the boundary conditions $\boldsymbol{n}\cdot\boldsymbol{\sigma}=\tilde{\boldsymbol{t}}$, $\boldsymbol{u}=\tilde{\boldsymbol{u}}$ and $\boldsymbol{n}\cdot\nabla d=0$ with $\bar{\boldsymbol{t}}$ and $\bar{\boldsymbol{u}}$ being the prescribed tractions and displacements on the corresponding boundaries, respectively.

To ensure crack irreversibility, in Eq.~(\ref{eq:goveq}), the crack driving force $\mathcal{H}$ in each point $\boldsymbol{x}$ is set to its temporal maximum \cite{miehe_phase_2016} 
\begin{equation} \label{eq:irr}
    \mathcal{H}(\boldsymbol{x},t) = \max_{s\in[t_0;t]} \psi^\mathrm{e}(\boldsymbol{\varepsilon}(\boldsymbol{x},s)).
\end{equation}

\subsection{Extension to fatigue}
\label{sec:model_Fat}

\subsubsection*{Fatigue degradation} 

In order to incorporate fatigue into the phase-field framework, the fracture toughness $\mathcal{G}_\mathrm{c}$ is reduced when the material degradation due to repetitive stressing precedes. This process is described by introducing a local lifetime variable $D$. An additional scalar fatigue degradation function \mbox{$\alpha(D):[0,1]\rightarrow[\alpha_0,1]$} with \mbox{$\alpha_0>0$} is introduced, which lowers the fracture toughness $\mathcal{G}_\mathrm{c}$ locally. The energy functional then reads
\begin{align}\label{eq:Pi_l}
\begin{split}
    \Pi_\ell =& \int_{\Omega} g(d)\,\psi^\mathrm{e}(\boldsymbol{\varepsilon}) \, \mathrm{d} V \\
    & + \int_{\Omega} \alpha(D) \,\mathcal{G}_\mathrm{c}\frac{1}{2\ell}(d^2+\ell^2|\nabla d|^2)\,\mathrm{d}V,
\end{split}
\end{align}
which leads to the modified phase-field evolution equation
\begin{equation}\label{eq:ev_PF}
   \mathcal{G}_\mathrm{c}\left( \alpha\,d-\nabla\alpha\cdot\ell^2\nabla d-\alpha\,\ell^2\Delta d\right)=(1-d)\,\mathcal{H}\,2\ell
\end{equation}
with the dependency $\alpha(D)$ dropped for brevity.

The lifetime variable $D\in[0,1]$ is a history variable that is accumulated strictly locally. For $D=0$, the material has experienced no cyclic loads and therefore offers full fracture toughness. Consequently, $\alpha(0)=1$ must hold. For $D=1$, the fracture toughness is reduced to a threshold value $\alpha_0$.      Therefore, the fatigue degradation function
\begin{equation}
    \label{eq:alpha}
    \alpha(D)=(1-\alpha_0)(1-D)^\xi+\alpha_0
\end{equation}
with the parameters $\alpha_0$ and $\xi$ is used. For a study of the influence of the model parameters $\alpha_0$ and $\xi$, see Sec.~\ref{sec:Sim_FCG} and \cite{seiler_efficient_2020}. 

\subsubsection*{Local strain approach}

The computation of the lifetime variable $D$ follows the LSA \cite{seeger_grundlagen_1996}. This method is generally used for fatigue life calculations of components, but is implemented here in the material routine of the FEM framework and therefore executed at each integration point. 

The computation scheme is illustrated in Fig.~\ref{fig:LSA}.
At first, the stresses and strains from a linear elastic simulation are revaluated using the Neuber rule \cite{neuber_theory_1961}. The von Mises equivalent stress $\sigma$ is projected to the cyclic stress-strain curve (CSSC) yielding a virtual, revaluated stress-strain pair $(\sigma^*,\varepsilon^*)$ by assuming a constant strain energy $\frac{1}{2}\,\sigma\varepsilon = \frac{1}{2}\,\sigma^*\varepsilon^*$. The CSSC is thereby described by the Ramberg-Osgood equation \cite{ramberg_description_1943}
\begin{equation}
    \varepsilon^* = \frac{\sigma^*}{E} + \left( \frac{\sigma^*}{K'} \right)^{1/n'}
\end{equation}
with the cyclic parameters $K'$ and $n'$ and Young's modulus $E$. It can be determined from standardised cyclic experiments such as the incremental step test. In this way, the complete virtual stress-strain path can be derived from the loading sequence. This stress-strain path is divided into hysteresis loops. For each loop $i$, the damage parameter by Smith, Watson and Topper \cite{smith_stressstrain_1970}
\begin{equation}
    \label{eq:PSWT}
    P_{\mathrm{SWT},i} = \sqrt{(\sigma_{\mathrm{a},i}^*+\sigma_{\mathrm{m},i}^*)\varepsilon_{\mathrm{a},i}^*E}
\end{equation}
can be determined from the stress and strain amplitudes $\sigma_\mathrm{a}^*$ and $\varepsilon_\mathrm{a}^*$ and the mean stress $\sigma_\mathrm{m}^*$. It quantifies the damaging effect of the loop. Only the tensile range contributes to $P_\mathrm{SWT}$, see~Fig.~\ref{fig:LSA}. From strain Wöhler curves (SWC) -- also generated with standardised experiments -- the matching virtual load cycle number $N_i$ for $P_{\mathrm{SWT},i}$ can be read. Finally, the fatigue life contribution of the single hysteresis loop $\Delta D_i$ and the full loading path is
\begin{equation}
    \label{eq:Miner}
    \Delta D_i = 1/N_i \quad\text{and}\quad D=\Sigma_i \Delta D_i.
\end{equation}
Note, that the revaluated stresses and strains $\sigma^*$ and $\varepsilon^*$ are solely used for the damage calculation and are not used in the coupled problem in any other way.

In conclusion, the integration of fatigue in the phase-field model with the LSA is beneficial for the computational cost in three ways:
\begin{enumerate}
    \item Local cyclic plasticity is covered by the Neuber rule, so no elastic-plastic material model is needed.
    \item Since only amplitude and mean values of stress and strain enter the calculation of $D$, the loading path does not have to be resolved in the simulation, instead the reversal points are sufficient.
    \item In case of constant load amplitudes and small crack growth rates, several load cycles can be simulated with only one increment, since the lifetime contributions are accumulated linearly according to Eq.~\ref{eq:Miner}. Especially for high cycle fatigue, this can save immense computational time.
\end{enumerate}

\subsection{Incorporation of residual stresses}
\label{sec:model_RS}

\begin{figure}  
    \def\svgwidth{\linewidth}   
    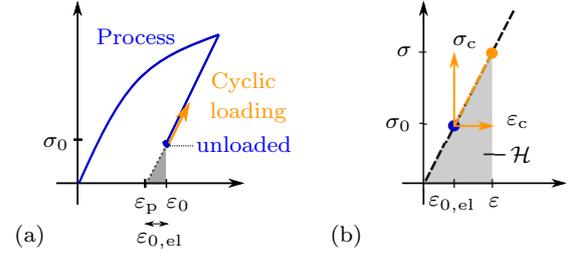
    \caption{\textbf{(a)} Formation of residual stresses $\sigma_0$ through plastic deformation. Remaining strain after unloading is $\varepsilon_0=\varepsilon_\mathrm{p}+\varepsilon_{0,\mathrm{el}}$. 
    \textbf{(b)} Material law for cyclic simulation. Initial state $(\sigma_0,\varepsilon_{0,\mathrm{el}})$ undergoes cyclic load with $(\sigma_\mathrm{c},\varepsilon_\mathrm{c})$. Crack driving force $\mathcal{H}$ is strain energy density of total stress-strain state $(\sigma,\varepsilon)$.
    Schemes of 1D, undamaged ($d=0$) case.}
    \label{fig:RS}       
\end{figure}

Residual stresses result from plastic deformations which occur during the production process, e.\,g. due to forming, tempering or surface treatment. The stress remaining after unloading is the residual stress $\boldsymbol{\sigma}_0$. The associated strain 
\begin{equation}
    \boldsymbol{\varepsilon}_0 = \boldsymbol{\varepsilon}_{0,\mathrm{el}} + \boldsymbol{\varepsilon}_\mathrm{p}
\end{equation}
consists of an elastic part $\boldsymbol{\varepsilon}_{0,\mathrm{el}}$ and a plastic part $\boldsymbol{\varepsilon}_\mathrm{p}$, see Fig.~\ref{fig:RS}(a). While the total residual strain $\boldsymbol{\varepsilon}_0$ is geometrically compatible, this does not apply to its components $\boldsymbol{\varepsilon}_\mathrm{p}$ and $\boldsymbol{\varepsilon}_{0,\mathrm{el}}$. 

Only the elastic part $\boldsymbol{\varepsilon}_{0,\mathrm{el}}$ of the residual strain $\boldsymbol{\varepsilon}_0$ is relevant for the fatigue life simulation. The plastic forming process is treated as completed and is not modelled. The plastic part $\boldsymbol{\varepsilon}_\mathrm{p}$ is not of further interest in the fatigue crack simulation, because it is assumed that the yielding process does not change the crack resistance properties of the material \cite{zerbst2016fatigue}. All material points are assigned the same material parameters initially, regardless of their (plastic) history.
Therefore the total stress-strain state $(\boldsymbol{\sigma},\boldsymbol{\varepsilon})$ in the model is the sum of the initial state $(\boldsymbol{\sigma}_0,\boldsymbol{\varepsilon}_{0,\mathrm{el}})$ and the stress-strain state caused by the cyclic loading $(\boldsymbol{\sigma}_\mathrm{c},\boldsymbol{\varepsilon}_\mathrm{c})$.
Hence, the total strain is
\begin{equation}
    \boldsymbol{\varepsilon} = \boldsymbol{\varepsilon}_{0,\mathrm{e}} + \boldsymbol{\varepsilon}_\mathrm{c}
\end{equation}
as displayed in Fig.~\ref{fig:RS}(b). Thereby, the strain $\boldsymbol{\varepsilon}_\mathrm{c}$ is
\begin{equation}
    \boldsymbol{\varepsilon}_\mathrm{c} = \frac{1}{2}\left(\nabla \boldsymbol{u} + (\nabla \boldsymbol{u})^\top\right).
\end{equation}
The regularised energy functional is, analogously to \linebreak Eq.~(\ref{eq:Pi_l}),
\begin{align}\label{eq:Pi_l_res}
\begin{split}
    \Pi_\ell =& \int_{\Omega} g(d)\,\psi^\mathrm{e}(\boldsymbol{\varepsilon}_{0,\mathrm{e}} + \boldsymbol{\varepsilon}_\mathrm{c}) \, \mathrm{d} V + \\ 
    & + \int_{\Omega} \alpha(D) \,\mathcal{G}_\mathrm{c}\frac{1}{2\ell}(d^2+\ell^2|\nabla d|^2)\,\mathrm{d}V.
\end{split}
\end{align}
Consequentially, the stress is 
\begin{equation}
    \boldsymbol{\sigma} = g(d) \left(\boldsymbol{\sigma}_0 + \boldsymbol{\sigma}_\mathrm{c} \right) = g(d) \left(\boldsymbol{\sigma}_0 + \mathbb{C}:\boldsymbol{\varepsilon}_\mathrm{c} \label{eq:stress_RS} \right).
\end{equation}
and the evolution equation remains
\begin{equation}\label{eq:ev_PF_res}
   \mathcal{G}_\mathrm{c}\left( \alpha\,d-\nabla\alpha\cdot\ell^2\nabla d-\alpha\,\ell^2\Delta d\right)=(1-d)\,\mathcal{H}\,2\ell.
\end{equation}
The crack driving force is the temporal maximum of the strain energy density of the total stress strain state
\begin{align}
    \mathcal{H}(t) &= \max_{s\in[t_0,t]} \left(\psi^\mathrm{e}(\boldsymbol{\varepsilon}_{0,\mathrm{el}} + \boldsymbol{\varepsilon}_\mathrm{c}(s)) \right)\\
    &= \max_{s\in[t_0,t]}\left( \frac{1}{2} ( \boldsymbol{\varepsilon}_{0,\mathrm{el}} + \boldsymbol{\varepsilon}_\mathrm{c}(s)) : \mathbb{C} : (\boldsymbol{\varepsilon}_{0,\mathrm{el}} + \boldsymbol{\varepsilon}_\mathrm{c}(s)) \right).
\end{align}
The initial state at the time $t_0$ is hereby $\boldsymbol{\varepsilon} = \boldsymbol{\varepsilon}_{0,\mathrm{el}}$, $\boldsymbol{\sigma}=g(d)\,\boldsymbol{\sigma}_0$.

The superposition of residual stress state and the stress state caused by external loading is actually also common in fracture-mechanical fatigue computations based on stress intensity factors \cite{larue_predicting_2007}. Note that since $\boldsymbol{\varepsilon}_{0,\mathrm{el}}$ is not necessarily geometrically compatible, the total strain $\boldsymbol{\varepsilon}$ is not, either. 

\begin{figure}
    \def\svgwidth{\linewidth}   
    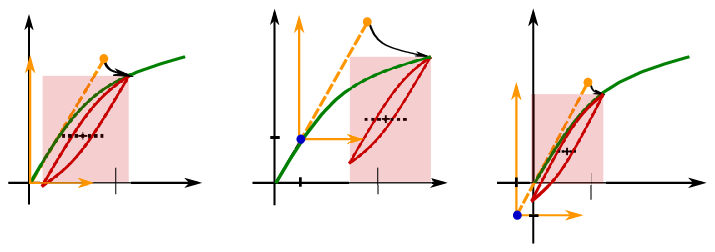
    \caption{Comparison of virtual stress-strain path with and without residual stress $\sigma_0$. Constitutive stress-strain behaviour (\protect\tikz[baseline=-0.5ex]{\protect\draw[thick,dashed] (0,0) -- (0.5,0) ;}) for 1D, undamaged ($d=0$) case. For the computation of the lifetime variable $D$, the virtual stress-strain path (\protect\tikz[baseline=-0.5ex]{\protect\draw[thick,color=darkred] (0,0) -- (0.5,0) ;}) is used, which is determined from the CSSC (\protect\tikz[baseline=-0.5ex]{\protect\draw[thick,color=darkgreen] (0,0) -- (0.5,0) ;}). Mainly, the residual stress shifts the virtual mean stress $\sigma_\mathrm{m}^*$, which controls the damage parameter $P_\mathrm{SWT}$ (schematically).}
    \label{fig:mat_RS}       
\end{figure}


It is assumed that the plastic strains do not enter the crack driving force. Instead the crack is driven by the elastic strain energy density, which again only depends on the total elastic strain. This assumption is appropriate for typical HCF and higher LCF loads which do not exceed the static yield limit.

Fig.~\ref{fig:mat_RS} depicts how the residual stresses affect the LSA procedure: Due to the initial state, the Neuber rule yields a shifted stress-strain path. Although the stress and strain amplitude stay the same, the damage parameter $P_\mathrm{SWT}$ is affected by the altered mean stress $\sigma_\mathrm{m}^*$ according to Eq.~\ref{eq:PSWT}. In this way, tensile residual stresses increase the damage parameter $P_\mathrm{SWT}$ while compressive residual lead to a decrease.

In summary, residual stresses influence the crack development in two ways:
\begin{enumerate}
    \item They change the peak stress-strain state of a load cycle which is decisive for the crack development in a cyclic load. In this way, tensile residual stresses increase, compressive residual stresses decrease the crack driving force $\mathcal{H}$ which depends on the strain energy density $\psi^\mathrm{e}(\boldsymbol{\varepsilon})$. 
    \item They shift the virtual stress-strain path and therefore influence the damaging effect and with that the lifetime variable $D$. Compressive residual stresses reduce $D$, while tensile residual stresses increase it.
\end{enumerate}

The initial residual stress tensor $\boldsymbol{\sigma}_0$ remains unchanged throughout the simulation. Residual stress redistribution due to wide-ranging plasticising is therefore not covered in the model, since this is only relevant for very low cycle fatigue with macroscopic plastic deformations \cite{benedetti_numerical_2010}. However, as cracks propagate, the stress state is rearranged due to degradation of the total stress (Eq.~\ref{eq:stress_RS}) which also affects the residual stresses. In this way, the residual stress state redistributes compared to the initial state $\boldsymbol{\sigma}_0$ due to FCG. 

\section{Experiments}
\label{sec:Exp}
The fatigue crack growth influenced by residual stresses is investigated with compact tension (C(T)) specimens, where significant residual stresses are introduced by LSP. The material under investigation is the aluminium alloy AA2024 in T3 heat treatment condition with 2\,mm and 4.8\,mm thickness, which is a representative aluminium alloy used in the aircraft industry for fuselage structures \cite{dursun2014recent}. The previous investigation \cite{keller_experimentally_2019} indicates that LSP allows the introduction of relatively high and deep residual stresses, where the microstructural changes in AA2024 does not influence the FCG behaviour significantly. Thus, differences in the FCG rates between untreated and laser peened material are mainly linked to the effect of residual stresses. In the following the LSP treatment, the experimental residual stress determination and the determination of the fatigue crack growth rate are described. The experimental data for specimens with thickness of 4.8\,mm are taken from \cite{kallien2019effect} and \cite{keller_experimentally_2019}.

\subsection{Laser shock peening}
\label{sec:Exp_LSP}
\begin{figure}
    \includegraphics[width=\linewidth]{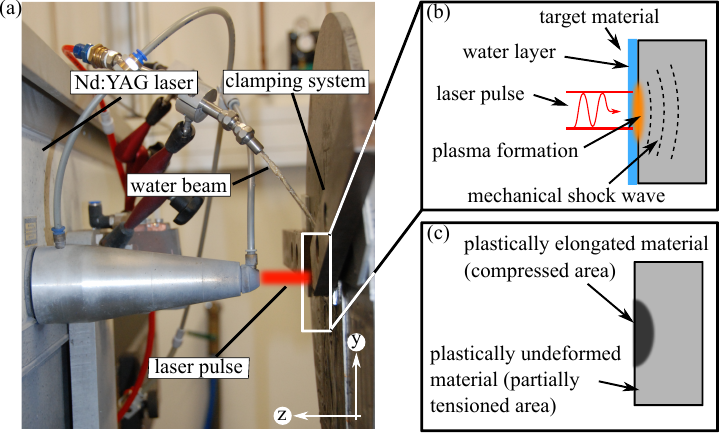}
    \caption{LSP laser system containing the pulsed Nd:YAG laser and clamping system fixing the specimen. Laser pulses are used to vaporize near-surface material to initiate mechanical shock waves \textbf{(b)}. These shock waves cause local plastic deformations, which lead to a residual stress distribution after the process \textbf{(c)}. A water layer increases the efficiency of the process by confining the plasma.  }
    \label{fig:LSP_Schema}       
\end{figure}

\begin{figure}
    \includegraphics[width=\linewidth]{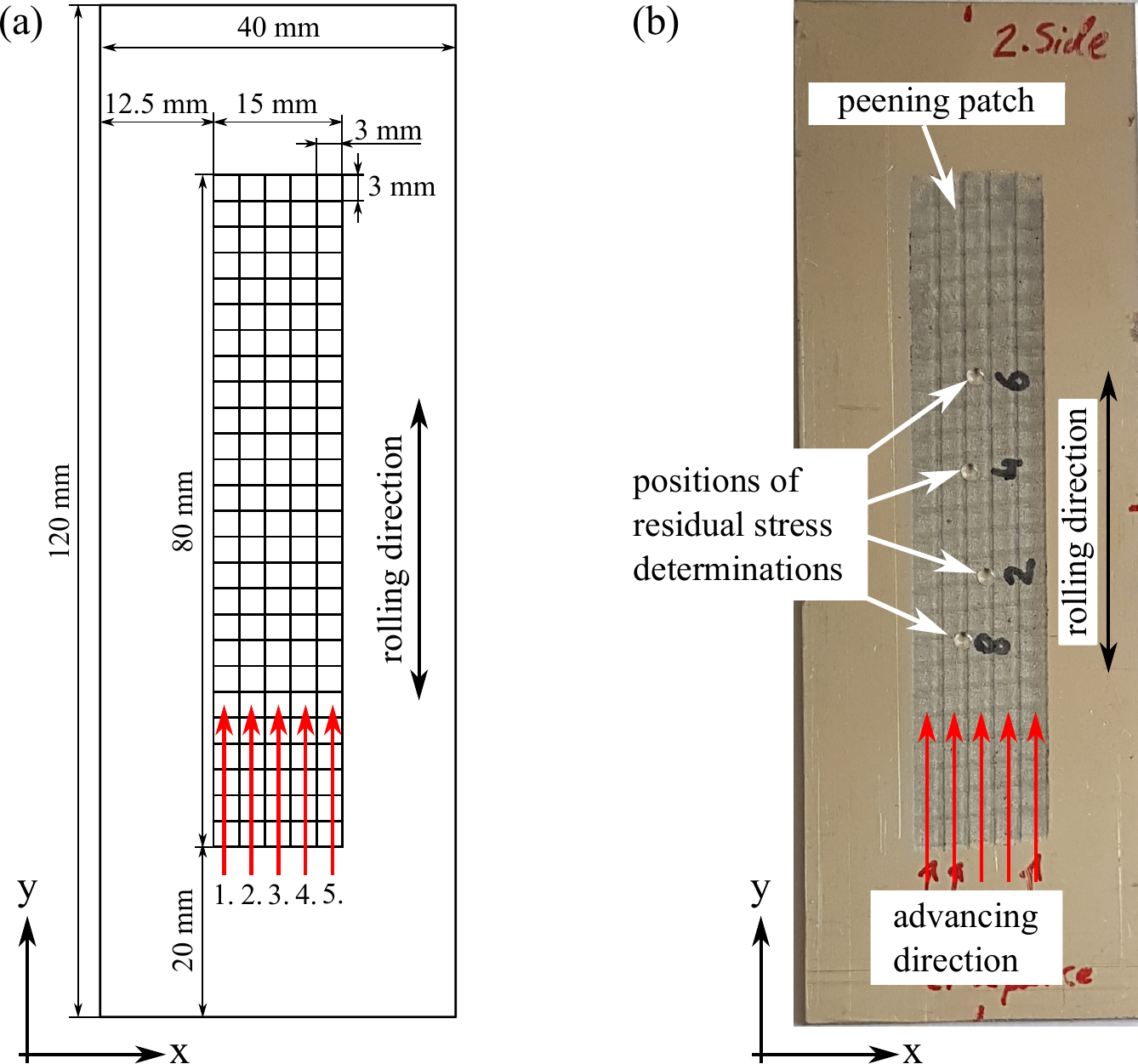}
    \caption{Schematic of the specimens used to evaluate the LSP-induced residual stress state \textbf{(a)}. 5\,J laser pulses with a $3 \, \mathrm{mm} \times 3 \, \mathrm{mm}$ square laser focus are used to treat a region with size $15 \, \mathrm{mm} \times 80 \, \mathrm{mm}$. The laser pulse sequence consists of five columns in which the laser pulses are shot without overlap. The advancing direction of each column is kept constant. The sequence is applied twice at both sides of the sheet material. Residual stresses were experimentally determined by incremental hole drilling \textbf{(b)}, where the same depth profile of residual stresses is assumed below the peening patch.}
    \label{fig:RS_Specimen}       
\end{figure}

LSP uses short-time high-energy laser pulses to generate plasma consisting of near-surface material, see Fig.~\ref{fig:LSP_Schema}. The extension of the plasma generates mechanical shock waves, which cause local plastic deformation of the subsurface material, see Fig.~\ref{fig:LSP_Schema}(b). After relaxation of the high dynamic process, these local plastic deformation lead to a residual stress distribution, where compressive residual stresses remain in the subsurface region surrounded by balancing tensile residual stresses, Fig.~\ref{fig:LSP_Schema}(c). The penetration depth of these compressive residual stresses is in millimeter range. The efficiency of the process can be increased by the use of a confinement layer. A laminar water layer is used during the LSP process in this study as confinement medium. The LSP treatment is conducted with an Nd:YAG laser. 5\,J laser pulses with the duration of 20\,ns (full width at half maximum) and a $3 \,\mathrm{mm} \times 3\,\mathrm{mm} $ square focus are used. The LSP treatment is performed without pulse overlap in five columns on the sheet material, as shown in Fig.~\ref{fig:RS_Specimen}(a). The laser pulse sequence of a rectangular peening patch with $15\, \mathrm{mm} \times 80\,\mathrm{mm}$ 
is applied twice at both sides of the sheet material. The sequence is shot at the first side twice before the second side was treated twice as well.

\subsection{Experimental residual stress determination}
\label{sec:Exp_RS}
The incremental hole drilling system PRISM from \linebreak Stresstech is used to determine the depth profile of the residual stresses. The hole drilling system uses electronic speckle pattern interferometry to determine material surface deformation after each increment of an incrementally drilled hole. These surface deformations are correlated with the residual stress at the respective increment depth via the integral method \cite{schajer2005full}. The interested reader is referred to \cite{ponslet2003residual,ponslet2003residual_PartIII,steinzig2003residual} for a detailed explanation of the incremental hole drilling method using electronic speckle pattern interferometry. 
A driller with 2\,mm diameter is used to determine the residual stresses up to the depth of 1\,mm. This hole depth allows for the experimental determination of the through thickness residual stress profile within the specimens with 2\,mm thickness, when the residual stresses are determined from both material sides. As it is recommended that the material thickness is four times larger than the hole diameter \cite{ponslet2003residual_PartIII}, the residual stress determinations of AA2024 with 2\,mm thickness where repeated with a hole diameter of 0.5\,mm as well. The determined residual stresses with the hole diameter of 0.5\,mm and 2\,mm match. Therefore, we focus on the determined residual stresses with the hole diameter of 2\,mm in the following. A relative small increment size is used near the material surface, where relatively large residual stress gradients are expected. Residual stresses were determined within the area of the peening patch, see Fig.~\ref{fig:RS_Specimen}(b). The depth profile of residual stresses is assumed to be the same below the whole peening patch. Thus, the average value and the standard deviation of at least eight experimentally determined residual stress profiles for both material sides are depicted in the following.

\subsection{Fatigue crack growth}
\label{sec:Exp_FCG}
\begin{figure}
    \includegraphics[width=\linewidth]{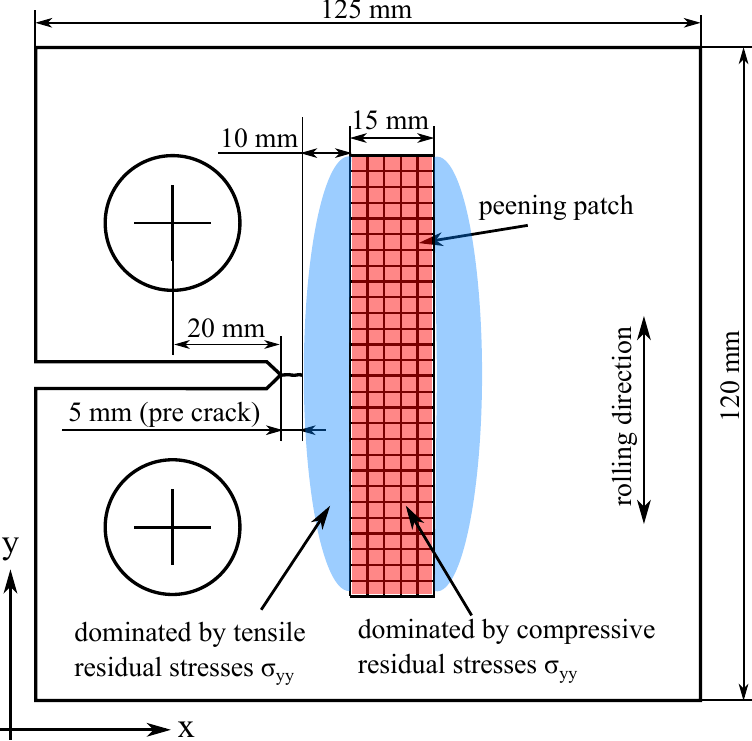}
    \caption{C(T) specimen with 100\,mm width according to ASTM-E647 standard. The specimen is pre-cracked by 5\,mm and subsequently peened with 10\,mm distance to the crack front of the pre-crack. C(T) specimens are treated twice from both surfaces of the sheet material. The introduction of compressive residual stresses below the peening patch lead to balancing tensile stresses in the surrounding material.}
    \label{fig:CT_Specimen}       
\end{figure}

C(T) specimens with a width of 100\,mm are used to determine the FCG rate according to ASTM~E647 standard. The FCG tests were performed with the servo hydraulically testing machine from Schenk/Instron and a 25 kN load cell. The specimen geometry is displayed in Fig.~\ref{fig:CT_Specimen}. A pre-crack of 5\,mm is introduced extending the initial crack length to 25\,mm (20\,mm notch and 5\,mm pre-crack). Afterwards, the peening patch, as described in Sec.~\ref{sec:Exp_LSP}, is applied 10\,mm in front of the initial crack front to the LSP treated specimens. The applied load ratio $R=0.1$ is kept constant during the FCG test. The maximum applied force of the fatigue load load cycles is 1.65\,kN and 4.0\,kN for specimens with 2\,mm and 4.8\,mm thickness, respectively. All FCG experiments are repeated at least twice.

\subsection{Experimental results}
\label{sec:Exp_result}

\subsubsection*{Residual stresses}
\begin{figure}  
    \def\svgwidth{\linewidth}   
    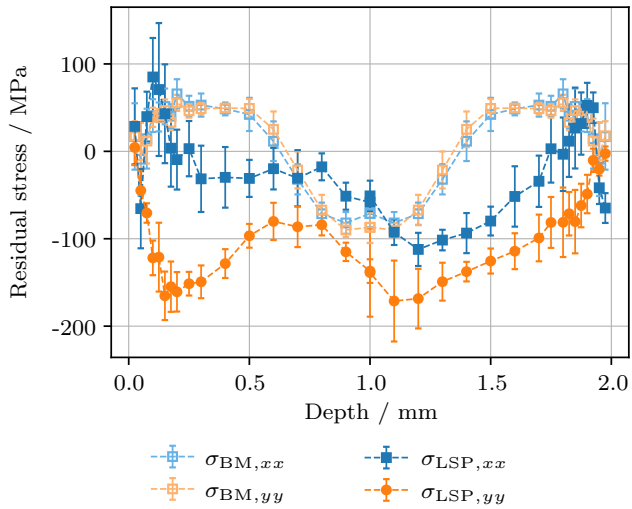
    \caption{Residual stress profile of the base material (BM) and after LSP in AA2024 with 2\,mm thickness. The average value and standard deviation are depicted. At least seven depth profiles were experimentally investigated. At first, LSP was applied at 0\,mm and secondly from the other side of the specimen at 2\,mm. The LSP treatment leads to compressive residual stresses along the entire material thickness.}
    \label{fig:RS_2mm}       
\end{figure}

\begin{figure} [h]
    \def\svgwidth{\linewidth}   
    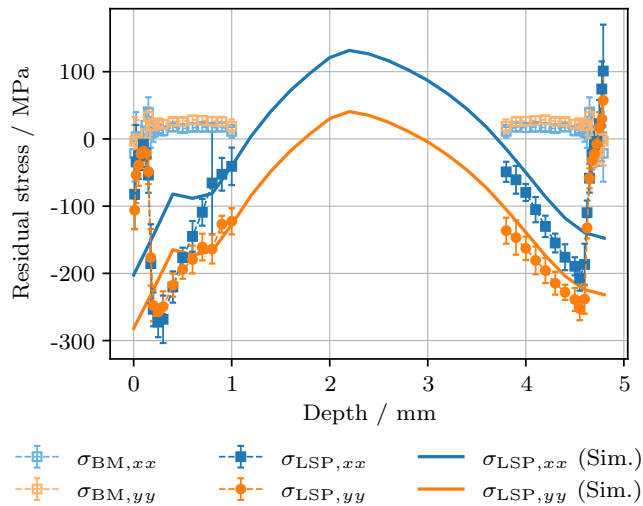
    \caption{Experimentally determined and numerically calculated (Sim.) residual stresses after LSP treatment in AA2024 with 4.8\,mm thickness. The firstly treated side is at 0\,mm. The numerically determined residual stress profile shows tensile residual stresses at mid-thickness (1.8-3.0\,mm). The residual stress profiles are taken from \cite{kallien2019effect} (BM) and \cite{keller_experimentally_2019} (LSP).}
    \label{fig:RS_4.8mm}       
\end{figure}

The LSP treatment leads to the introduction of significant compressive residual stresses over the thickness for the investigated AA2024, see Fig.~\ref{fig:RS_2mm} for 2\,mm thickness and Fig.~\ref{fig:RS_4.8mm} for 4.8\,mm thickness, respectively. 

Since the residual stresses are only determined up to a depth of 1\,mm from each side with the incremental hole drilling technique, numerical simulations via a LSP process simulation, as described elsewhere \cite{keller_crack_2019,keller_experimentally_2019} are used to estimate the residual stress profile along the entire material cross-section in $z$ direction.

The residual stress components $\sigma_{xx}$ and $\sigma_{yy}$ in surface plane direction differ, whereby the magnitude of the component perpendicular to the crack growth direction, $\sigma_{yy}$, is more pronounced. This difference of the residual stress components might be attributed to geometrical effects, such that the rectangular peening patch geometry, as experiments with a square peening patch do not indicate this significant difference of $\sigma_{xx}$ and $\sigma_{yy}$ in aluminium alloy AA2024 \cite{keller_experimentally_2019}. 

The residual stress magnitude and gradient differ significantly depending on the material thickness. The LSP-treated aluminium alloy with 2\,mm thickness contains a lower maximum compressive residual stress of approximately 160\,MPa compared to the compressive maximum of approx\-imate\-ly 280\,MPa in the 4.8\,mm thick material. While tensile residual stresses occur at mid-thickness for the thicker material, the residual stress component $\sigma_{yy}$ is completely compressive along the $z$ direction for the 2\,mm thick material. 

The resulting residual stress field depends on the order of the applied pulse sequences on the two sides. These differences are more pronounced for the thinner material. The non-symmetric residual stress profile is assumed to result from the interaction of mechanical shock waves initiated at the secondly peened side at $z=2 \, \mathrm{mm}$ and already existing residual stresses introduced by the LSP treatment of the first side at $z=0 \, \mathrm{mm}$. These interactions result in increased residual stresses between 0.4\,–\,0.9\,mm depth. 

It has to be noted, that the peening patch surrounding material in $x$-$y$ direction contains balancing tensile residual stresses. A detailed analysis about the overall residual stress field of the C(T) specimen after the LSP treatment in AA2024 for 4.8\,mm thickness can be found in \cite{keller_experimentally_2019}.

\subsubsection*{Fatigue crack growth} 
\begin{figure}  
    \def\svgwidth{\linewidth}   
    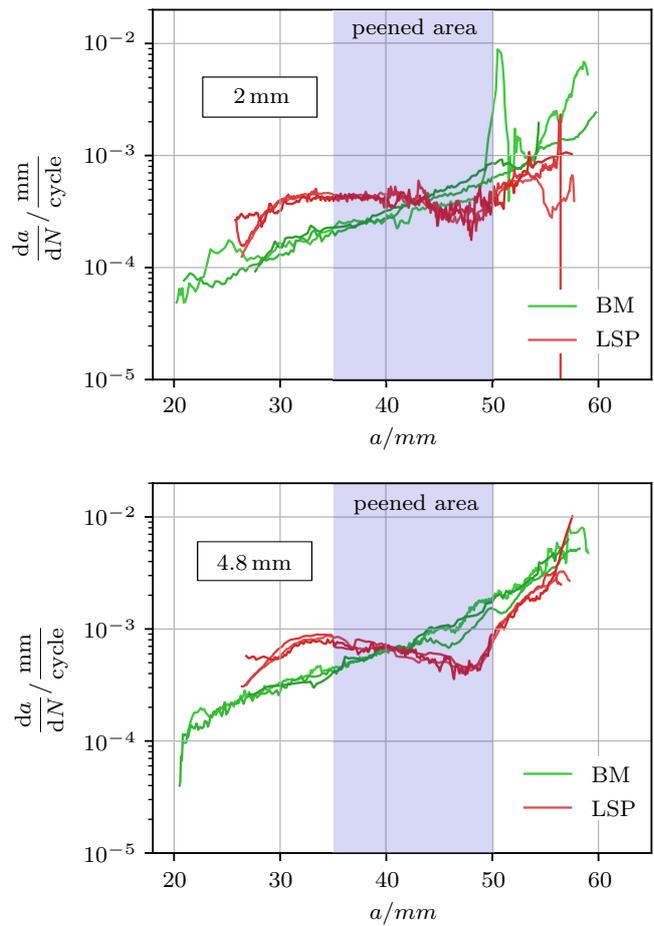
    \caption{FCG rate in AA2024 with 2\,mm and 4.8\,mm thickness in BM and LSP-treated material. The introduced residual stress field leads to an accelerated FCG in front of the peened area and a FCG retardation when the crack front is located within the peening patch. The different curves, represent the repeated experiments. The data for the 4.8\,mm specimen are taken from \cite{keller_experimentally_2019}}
    \label{fig:FCG_mm}       
\end{figure}

Experimentally determined FCG rates of the untreated material show the typical exponential correlation between FCG rate $\mathrm{d}a/\mathrm{d}N$ and stress intensity factor range $\Delta K$ known for the Paris regime, see Fig.~\ref{fig:FCG_mm}. This characteristic FCG behaviour is significantly affected by the introduced residual stresses for both investigated material thicknesses. For the LSP-treated samples, the FCG rate increases between the initial crack front and the peening patch. This increased FCG rate is attributed to balancing tensile residual stresses, as indicated in Fig.~\ref{fig:CT_Specimen}. Thereafter, the FCG rate decreases up to a minimum at $a \approx 49\,\mathrm{mm}$, when the crack front is located within the area of the peening patch. After the crack front has passed the peening patch, the FCG rate accelerates, but stays below the FCG rate of specimens without LSP treatment. This characteristic FCG behaviour is observed for both material thicknesses.

The observation of the increased FCG rate highlights the importance of the overall residual stress field for an efficient application of residual stress modification techniques, such as LSP. Furthermore, tools for FCG rate calculation need to contain the prediction of this possible increase of the FCG rate as well. While the material thickness of 2\,mm allows the experimental determination of residual stresses over entire material thickness, the relatively thin material may lead to buckling during the fatigue testing at larger crack length. These buckling phenomena are indicated from $a > 50 \, \mathrm{mm}$ and may cause the increased scatter of the experimentally determined FCG rate for the material thickness 2\,mm.

\section{Simulation of fatigue crack growth}
\label{sec:Sim}

In the following, the phase-field model described in Sec.~\ref{sec:model} is used to simulate the fatigue crack growth experiments described in Sec.~\ref{sec:Exp_FCG}. Starting with an unpeened specimen, model parameters are studied and the model is calibrated to one fatigue crack growth curve. With the calibrated model parameters, the other unpeened and peened specimens are simulated.

For all simulations, a staggered solution scheme \cite{hofacker_continuum_2012} is applied to solve the coupled problem with mechanical field and phase-field. A structured, locally refined mesh with a minimum mesh size of $h_{\min}=0.33$\,mm is used. Due to the thin specimens, a plane stress state is assumed. The plane stress assumption is supported by the experimental determination of the residual stress component perpendicular to the material surface ($\sigma_{zz} \approx 0\,\mathrm{mm}$) after LSP application via synchrotron radiation in \cite{keller2018experimental}. As comparative computations showed, a tension-compression split, see \cite{miehe_phase_2010}, is not necessary for the simulations due to the simple stress state of the specimen. The characteristic length is set to $\ell=1$\,mm as a compromise between mesh refinement and accuracy. The elastic, cyclic and fracture-mechanical parameters for the AA2024-T3 material taken from literature are specified in Tab.~\ref{tab:matpar}. 

The specimen are loaded by cyclic loading with force ratio $\tilde{F}_\mathrm{min}/\tilde{F}_\mathrm{max}=0.1$. The force boundary condition was kept to $\tilde{F}_\mathrm{max}$ throughout the simulation. This is possible due to the model formulation with the LSA. The damage parameter $P_\mathrm{SWT}$ only needs amplitude and mean stress values as an input, see Sec.~\ref{sec:model_Fat}. Therefore, a simulation of the full loading path is not necessary for the damage calculation.
Due to the constant amplitude loading, several load cycles $\Delta N$ can be simulated within one increment. $\Delta N$ is reduced adaptively depending on the number of Newton iterations required to in the staggered loop, starting with $\Delta N=3000$.

\begin{table}
    \caption{Material parameters of aluminium AA2024-T3 used in phase-field simulations.}
    \label{tab:matpar}       
    \begin{tabular}{lll}
    \hline\noalign{\smallskip}
    Elastic constants \cite{boller_materials_1987} & $E=74.6$ GPa & $\nu=0.33$ \\
    CSSC \cite{boller_materials_1987}  & $K'=0.453$ GPa & $n'=0.201$  \\
    SWC \cite{boller_materials_1987} & $\sigma'_\mathrm{f}=0.314$ GPa & $\varepsilon'_\mathrm{f}=0.162$  \\
    & $b=-0.091$ & $c=-0.452$ \\
    Fracture toughness \cite{kaufman_fracture_2001} & $\mathcal{G}_\mathrm{c}=0.165\, \mathrm{MPa\,m}$ & \\
    \noalign{\smallskip}\hline
    \end{tabular}
\end{table}

\subsection{Unpeened specimens}
\label{sec:Sim_FCG}

\subsubsection*{Model parameters} 

\begin{figure}
    \def\svgwidth{\linewidth}   
    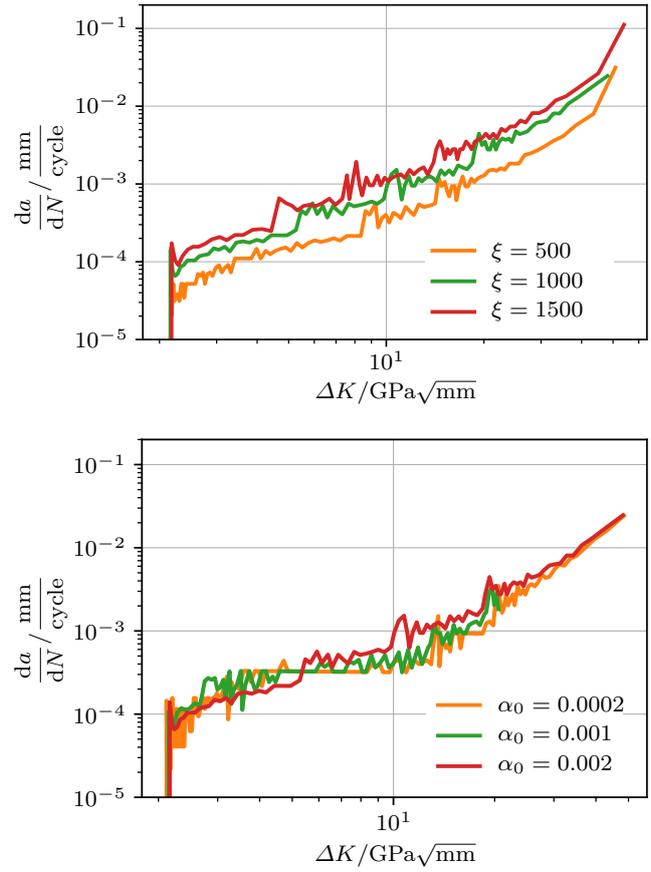
    \caption{FCG rate for range of stress intensity factor $\Delta K$ (Paris curves). Study of the exponent $\xi$ and the threshold $\alpha_0$ of the fatigue degradation function.}
    \label{fig:alpha_xi}       
\end{figure}

The parameters of the fatigue degradation function (\ref{eq:alpha}), $\alpha_0$ and $\xi$, are the only model parameters that have to be calibrated apart from the characteristic length $\ell$. All other parameters -- listed in Tab.~\ref{tab:matpar} -- are drawn from standardised experiments. The influence of the fatigue degradation function is studied on the 4.8\,mm thick specimen loaded with \mbox{$\tilde{F}_\mathrm{max}=4\,\mathrm{kN}$}. Fig.~\ref{fig:alpha_xi} shows the results as a Paris plot, i.\,e. the crack growth rate over the range of the stress intensity factor. The variation of the threshold value $\alpha_0$ while keeping $\xi=1000$ shows its influence on the inclination and curvature of the Paris curve. For \mbox{$\alpha_0=0.002$}, the graph is a straight line in the double logarithmic plot, which is typical for most crack growth experiments. Varying the exponent~$\xi$ while keeping $\alpha_0=0.002$ shifts the Paris curve in vertical direction. 
For a study of the influence of the model parameters on crack \textit{initiation}, the reader is referred to \cite{seiler_efficient_2020}.

\subsubsection*{Model calibration}

\begin{figure}
    \def\svgwidth{\linewidth}   
    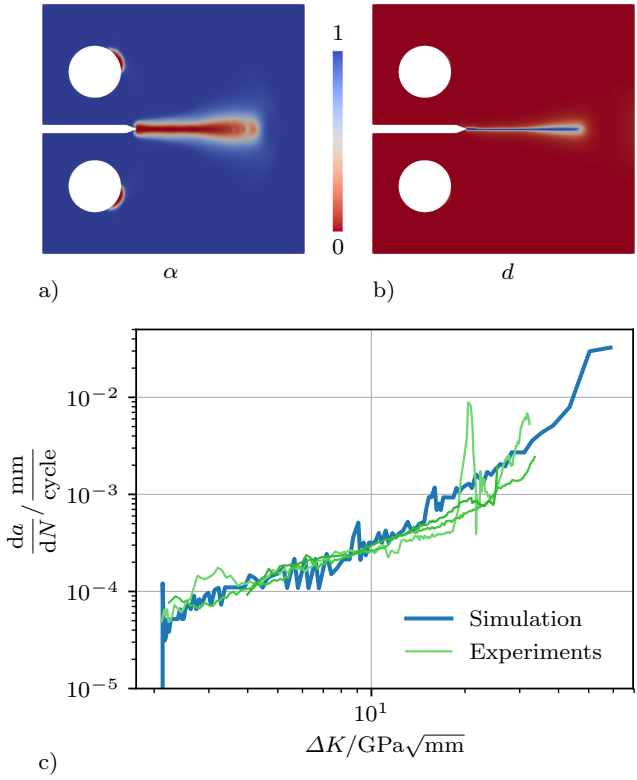
\caption{Fatigue crack growth of a 2\,mm thick specimen loaded with $\tilde{F}_\mathrm{max}=1.65\,\mathrm{kN}$. Model parameters are fitted to experimental results yielding $\alpha_0=0.0015$ and $\xi=500$. \textbf{a)} Fatigue degradation $\alpha$ and \textbf{b)} phase-field variable $d$ after $\approx251\,500$ load cycles. \textbf{c)} Paris curve.}
\label{fig:Exp488}       
\end{figure}

The different effects of the two parameters allow for a convenient calibration of the model. Here, a fatigue crack growth experiment with the 2\,mm thick specimen loaded with \mbox{$\tilde{F}_\mathrm{max}=1.65\,\mathrm{kN}$}, which  was repeated three times, is used for calibration. The fit of the Paris curve yielded the model parameters $\alpha_0=0.0015$ and $\xi=500$ as displayed in Fig.~\ref{fig:Exp488}. The figure also shows the distribution of the fatigue degradation $\alpha$ and the crack indicating phase-field variable $d$ after $251\,500$ load cycles.

\subsubsection*{Test loading case}
\begin{figure}
    \def\svgwidth{\linewidth}   
    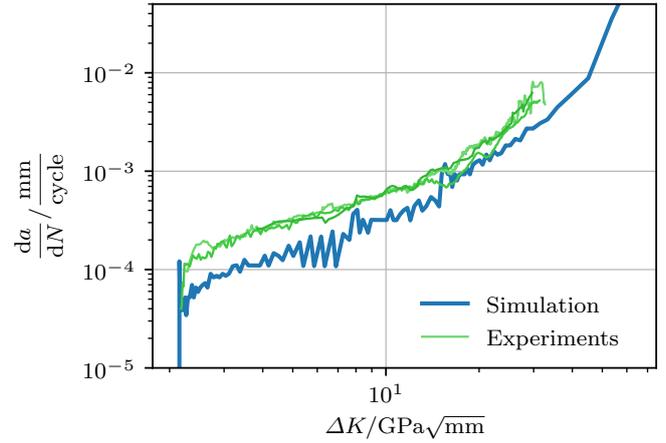
    \caption{Fatigue crack growth in a 4.8\,mm thick specimen loaded with $\tilde{F}_\mathrm{max}=4\,\mathrm{kN}$. Test loading case with calibrated model parameters $\alpha_0=0.0015$ and $\xi=500$.}
    \label{fig:Exp496}       
\end{figure}

The calibrated parameters are tested with a different a fatigue crack growth experiment with a 4.8\,mm thick specimen loaded with $\tilde{F}_\mathrm{max}=4\,\mathrm{kN}$, also repeated three times. As displayed in Fig.~\ref{fig:Exp496}, the simulation meets the experiments quite well, yet underestimates the crack growth rate slightly. This could be due to the fact that the assumption of a plane stress state as it was used in the simulation is less accurate for a thicker (albeit still thin) specimen. 

\subsection{Peened specimens}
\label{sec:Sim_RS}

\begin{figure}
    \def\svgwidth{\linewidth}   
    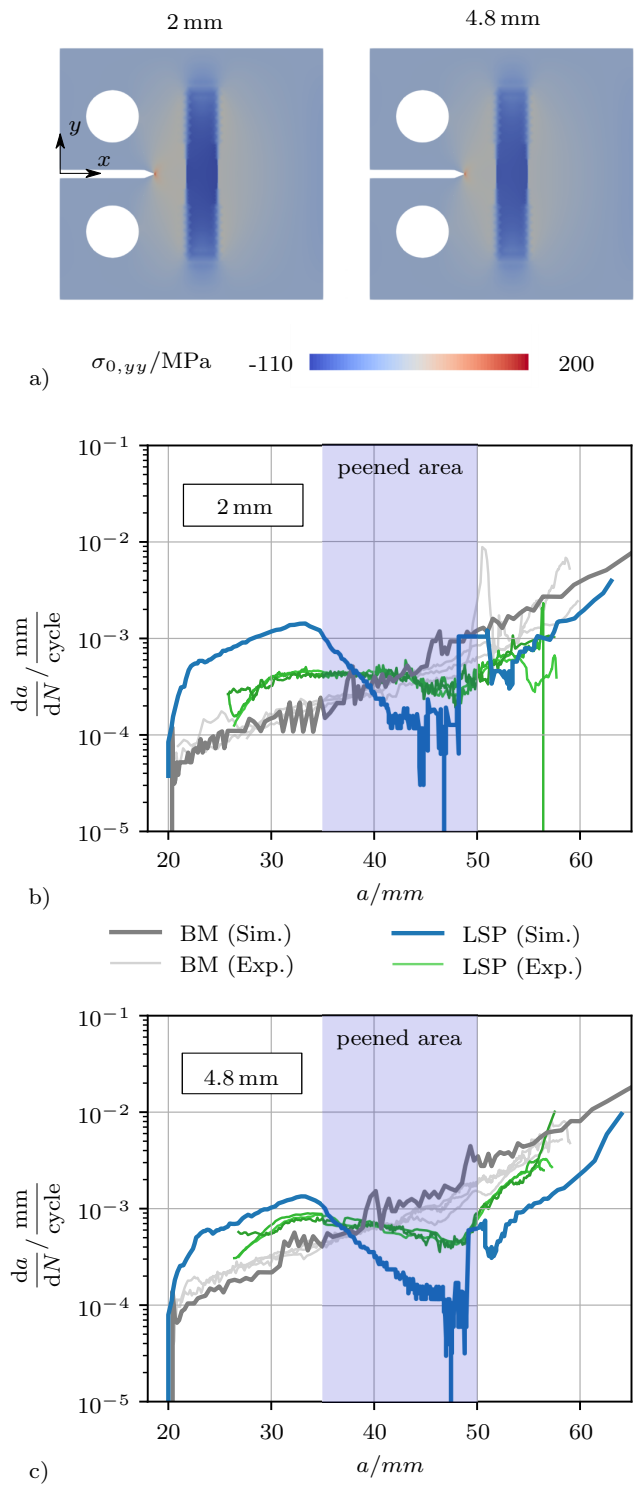
    \caption{Fatigue crack growth in 2\,mm and 4.8\,mm thick peened and unpeened specimen loaded with $\tilde{F}_\mathrm{max}=1.65\,\mathrm{kN}$ and $\tilde{F}_\mathrm{max}=4\,\mathrm{kN}$. Model parameters $\alpha_0=0.0015$ and $\xi=500$. \textbf{a)} Imposed residual stress component $\sigma_{0,yy}$ from residual stress measurements after LSP. The component $\sigma_{0,xx}$ is not shown here. \textbf{b)} and \textbf{c)} Crack growth rate.} 
    \label{fig:Comp1074_1077}       
\end{figure}

Before the crack growth simulation, the initial residual stress state has to be established. For this purpose, the experimentally determined residual stresses are mapped to the used mesh. In this context, the integral mean of the depth profile of the experimentally determined residual stresses $\sigma_{0,xx}$ and $\sigma_{0,yy}$ is taken to fit the 2D plane stress simulation. The integrated shear stress as well as the stress in thickness direction is close to zero. With a preliminary, load-free simulation, an equilibrium stress state is found which serves as the initial residual stress $\boldsymbol{\sigma}_0$ in the actual simulation. For both, the 2\,mm and the 4.8\,mm thick specimen, the employed residual stress component $\sigma_{0,yy}$ is depicted in Fig.~\ref{fig:Comp1074_1077}a) exemplarily. 

Both peened specimens are now simulated with the parameters fitted in the previous section. Please note that no additional parameters are modified for the simulations including the residual stresses. The initial load cycle increments are $\Delta N=300$ and $\Delta N=1000$, respectively. Fig.~\ref{fig:Comp1074_1077} shows the FCG rates. The peened area is shaded. Within this area, there are dominantly compressive residual stresses, while before and after it the residual stresses are primarily in the tensile range. Both simulations reproduce the effect of the peening very well qualitatively: The crack is accelerated in front and after the peened area while within the peened area it is inhibited. 

The model overestimates the influence of the residual stresses. This is true for both crack accelerating and inhibiting effects. One reason for the quantitative gap between experiment and simulation is that crack closure is not considered in the model. Another reason could be the simplification to a 2D stress state. The residual stresses introduced through LSP have a distinct profile over the thickness which influence the FCG rate. 

The oscillations at the end of the peened area result from the fact that the very low FCG rates in this area lead to almost zero crack growth in some increments. The jump at the end of the peened area presumeably stems from the high residual stress gradient applied in the simulation due to the fact that measurements of residual stresses are only pointwise.

\section{Conclusion}
\label{sec:Conc}

This paper revisits a phase-field model for the computationally effective simulation of fatigue cracks \cite{seiler_efficient_2020}. The model is calibrated and validated with FCG experiments in aluminium metal sheets. It is able to reproduce different FCG experiments fairly well. In the second part, residual stresses are introduced into the metal sheets through LSP. A method for the incorporation of residual stresses into the model is presented. The model is able to reproduce the crack inhibiting effect of the compressive residual stresses qualitatively.

Future works will now focus on low and very low cycle fatigue where the elastic approximation is not valid anymore. Moreover the degradation of residual stresses due to cyclic plasticity deserves closer attention. A 3D simulation which considers the distinct crack closure effects over the thickness of the specimen could also yield more realistic results.

\begin{acknowledgements}
The group of M. Kästner thanks the German Research Foundation DFG which supported this work within the Priority Programme 2013 ''Targeted Use of Forming Induced Residual Stresses in Metal Components'' with grant number KA 3309/7-1. The authors would like to thank M. Horstmann and H. Tek for the specimen preparation and performing the fatigue tests.
\end{acknowledgements}

\textbf{Conflict of interest} 
On behalf of all authors, the corresponding author states that there is no conflict of interest. \\

This is a preprint of an article published in \textit{Archive of Applied Mechanics}. The final authenticated version is available online at: https://doi.org/10.1007/s00419-021-01897-2.

%
%

\bibliographystyle{spmpsci}      
\bibliography{Literature.bib}   

%
%

\end{document}